\documentclass[nofootinbib,prl,twocolumn,showpacs,showkeys,preprintnumbers]{revtex4-1}
\usepackage{hyperref,amssymb,amsmath,mathrsfs,bm,graphicx,color}
\usepackage{lineno}
\begin{document}

\title {All analytic solutions for geodesic motion in axially symmetric space-times}
\author{ J. Ospino, J.L. Hern\'andez-Pastora}
\email{j.ospino@usal.es} \email{jlhp@usal.es}
\affiliation{Departamento de Matem\'atica Aplicada and Instituto Universitario de F\'isica
Fundamental y Matem\'aticas, Universidad de Salamanca, Salamanca 37007, Spain}

\author{and L.A. N\'u$\tilde{n}$ez}
\email{ lnunez@uis.edu.co}
\affiliation{ Escuela de F\'isica, Universidad Industrial de Santander, Bucaramanga, Colombia\\
 Departamento de F\'isica, Universidad de Los Andes, M\'erida, Venezuela}

\begin{abstract}
Recent observations of the orbits of star clusters around Sgr $A^\star$, imaging of black holes and gravitational waveforms of merging compact objects require a detailed understanding of the general relativistic geodesic motion. We came up with a method to provide all the possible geodesics in an axially symmetric space-time. The Kerr metric is explicitly worked out, recovering the Schwarzschild geodesics in the static limit. We also found the most general Killing tensor and its associated constant of motion for an axisymmetric space-time. The relevance of these results is crucial to understanding the different scenarios and the fundamental nature of the compact object at the galactic center.
\end{abstract}
\pacs{04.40.-b, 04.40.Nr, 04.40.Dg}
\keywords{ Nonspherical sources, Exterior solutions. Geodesic motions}
\maketitle

\section{Introduction}
 From the orbits of stars~\cite{GhezEtal2008, GenzelEisenhauerGillessen2010, AbuterEtal2018A}, and the imaging around supermassive black holes~\cite{AkiyamaEtal2019EHTColl, PsaltisEtalEHT2020} through gravitational lensing~\cite{HeLin2021}, geodesic motions of particles and photons have a venerable history bringing results of General Relativity to observational grounds. Other important scenarios like frame dragging\cite{CostaNatario2021},  radiation transport effects from accretion flows in the vicinity of relativistic stars~\cite{YuanNarayan2014}, and gravitational waveform of merging compact  objects~\cite{AbbottEtall2016LIGOVIRGOColl}, also require a detailed understanding of the geodesic motion in a general relativistic background.

Rotation is a crucial feature for celestial bodies from the an astrophysical viewpoint, and the Kerr solution of Einstein equations describes the gravitational field outside spinning compact objects and black holes~\cite{Kerr1963}. Thus, geodesics across a Kerr gravitational background are very important and have a long history, since Carter's study of the existence of a new conserved quantity associated with each geodesic~\cite{Carter1968, BCarter1968}. The astrophysical relevance of tracking particles \& photons in Kerr space-times motivates a significant effort to obtain analytical and numerical trajectories in this gravitation background (see \cite{Teo2021, ChanEtal2018} and references therein).

We have recently implemented a tetrad formalism by an orthogonal splitting of the Riemann tensor, introducing a complete set of equations equivalent to the Einstein system and applying it to the spherical case~\cite{OspinoHernandezNunez2017, OspinoEtal2018, OspinoNunez2020}. This formalism provides coordinate-free results expressed in terms of structure scalars related to the kinematical and physical properties of the fluid.

We devise an exhaustive classification for all geodesic motion for any axisymmetric source establishing the equations which describe each alternative. Based on the tetrad scheme, we provide a method to integrate all possible orbits for any stationary axially symmetric solutions,  illustrating each case with the Kerr metric.

We present the most general Killing tensor corresponding to any axisymmetric space-time and its associated constant of motion, which has not previously been obtained. This Killing tensor and its conserved quantity allow us to obtain the orbits by solving a system of algebraic equations. We recovered the famous Carter constant along the geodesic as a particular case.

\section{The tetrad \& kinematical variables}
We shall consider stationary and axially symmetric sources with the line element written as
\begin{equation}
{\rm d}s^2=-A^2 {\rm d}t^2 +B^2{\rm d}r^2 +C^2{\rm d}\theta^2 +R^2 {\rm d}\phi^2 +2\omega_3 {\rm d}t \, {\rm d}\phi \, ,
\label{Axisymmetric}
\end{equation}
with $A = A( r, \theta)$, $B = B( r, \theta)$, $R = R( r, \theta)$, and $\omega_3~=~\omega_3( r, \theta)$.

In this case the tetrad is:
\begin{eqnarray}
  V^\alpha&=&\left(\frac{1}{A},0,0,0\right), \quad
  K^\alpha = \left(0,\frac{1}{B},0,0\right),\nonumber  \\
   \quad
   L^\alpha & = & \left(0,0,\frac{1}{C},0\right),   S^\alpha=\frac{1}{\sqrt{\Delta_2}}\left(\frac{\omega_3}{A},0,0,A\right)\, ,\nonumber
\end{eqnarray}
where  $\Delta_2=A^2R^2+\omega_3^2$.

\subsection{The scalars and the tetrad covariant derivative}
The covariant derivative of $V_{\alpha}$ in the 1+3 formalism can be written as
$V_{\alpha;\beta}=-a_\alpha V_\beta+\Omega_{\alpha\beta}$,
where  the kinematical variables ($a_\alpha$ the acceleration and  $\Omega_{\alpha\beta}$ the vorticity) can be written, in terms of the tetrad, as
\begin{eqnarray}
a_\alpha&=&a_1K_\alpha +a_2L_\alpha\, , \\
 \Omega_{\alpha \beta}&=&\Omega_2(K_\alpha S_\beta-K_\beta S_\alpha)+\Omega_3 (L_\alpha S_\beta-L_\beta S_\alpha) \, .
\end{eqnarray}

Follows the covariant derivative of {\bf K}, i.e.
\begin{eqnarray}
K_{\alpha;\beta}&=&-a_1 V_\alpha V_\beta+2\Omega_2 V_{(\alpha}S_{\beta)}\nonumber\\
&+&(j_1K_\beta+j_2L_\beta)L_\alpha+j_6 S_\alpha S_\beta \, .\nonumber
\end{eqnarray}

Now, the covariant derivative of {\bf L}, can be written as
\begin{eqnarray}
L_{\alpha;\beta}&=&-a_2 V_\alpha V_\beta+2\Omega_3 V_{(\alpha}S_{\beta)}\nonumber\\
&-&(j_1K_\beta+j_2L_\beta)K_\alpha+j_9 S_\alpha S_\beta \, .\nonumber
\end{eqnarray}

Finally the covariant derivative of {\bf S} is
\begin{equation}
S_{\alpha;\beta}=-2\Omega_2 V_{(\alpha}K_{\beta)}-2\Omega_3 V_{(\alpha}L_{\beta)}-(j_6K_\alpha+j_9L_\alpha)S_\beta \, . \nonumber
\end{equation}

\subsection{Scalars for a general axisymetric  metric}
Assuming $\omega_3=A^2 \psi$, the scalars for the axisymmetric metric (\ref{Axisymmetric})  are:
\begin{eqnarray}
  a_1 &=&\frac{A_{,r}}{AB} \, , \qquad a_2= \frac{A_{,\theta}}{AC} \, , \\
   j_1 &=&-\frac{B_{,\theta}}{BC} \, , \qquad j_2= \frac{C_{,r}}{BC} \, , \\
  j_6 &=& \frac{(\ln (R^2+A^2 \psi^2))_{,r}}{2B} \, , \\
  j_9 &=& \frac{(\ln (R^2+A^2 \psi^2))_{,\theta}}{2C} \, , \\
  \Omega _2&=& \frac{A\psi _{,r}}{2B\sqrt{R^2+A^2\psi^2}} \quad {\rm and}\\
\Omega _3&=& \frac{A\psi _{,\theta}}{2C\sqrt{R^2+A^2\psi^2}}\, .
\end{eqnarray}

\section{All Geodesic}
To obtain all possible geodesics in any axially symmetric space-time, we define the tangent vector to the geodesic $Z^{\alpha}$ as
  \begin{equation}
  Z^\alpha \equiv \frac{{\rm d}x^\alpha}{{\rm d}\lambda} = z_0 V^\alpha +z_1K^\alpha  +z_2L^\alpha +z_3S^\alpha \, ,
  \end{equation}
and, its norm, $ \epsilon = Z_\alpha Z^\alpha = -z_0^2 +z_1^2  +z_2^2 +z_3^2$,
represents photon ($\epsilon = 0$) and particle ($\epsilon = -1$) trajectories.

In what follows we shall make use of the geodesic equations $Z_{\alpha;\beta}Z^\beta=0$, written in the tetrad formalism as
\begin{equation}
        z_1 z_1^\dag+z_2z_1^\theta = j_1z_1 z_2+2z_0z_3 \Omega_2-a_1z_0^2 +j_2 z_2^2 +j_6z_3^2 \label{G1}
\end{equation}
and
\begin{equation}
        z_1z_2^\dag+z_2z_2^\theta = -j_2z_1z_2+2z_0z_3\Omega_3-a_2z_0^2-j_1 z_1^2+j_9z_3^2 \label{G2} \, .
\end{equation}

Because the norm of the tangent vector $Z^\alpha$ is constant: $Z_\alpha Z^\alpha =\epsilon \Rightarrow\quad Z_{\alpha;\beta} Z^\alpha=0$, and we get
\begin{eqnarray}
    z_1 z_1^\dag+z_2 z_2^\dag&=&j_6z_3^2+2z_0z_3\Omega_2-a_1z_0^2 \quad {\rm and} \label{M2}  \\
     z_1 z_1^\theta+z_2 z_2^\theta&=&j_9z_3^2+2z_0z_3\Omega_3-a_2z_0^2 \, ; \label{M3}
\end{eqnarray}
where we have the derivatives $z^\dag \equiv\frac{1}{B}z_{,r} $ and    $z^\theta \equiv \frac{1}{C}z_{,\theta}$.

\subsection{First order geodesic equations}
Substracting (\ref{M2}) from (\ref{G1}) and (\ref{M3}) from (\ref{G2}) we have
\begin{eqnarray}
  z_2(z_2^\dag-z_1^\theta+j_1z_1+j_2 z_2)&=&0 \quad {\rm and} \label{M2G1}\\
  z_1(z_2^\dag-z_1^\theta+j_1z_1+j_2 z_2)&=&0 \, . \label{M3G2}
\end{eqnarray}
This system describes all possible geodesic equations for any axially symmetric space-time, having Kerr and Schwarzschild metrics as particular cases.

\subsection{Solutions for all geodesic cases}
The four cases which solve the system (\ref{M2G1})-(\ref{M3G2}):
\begin{enumerate}
	\item $z_2=0$ and $z_1=0$, representing circular orbits on constant planes, $\theta = const$.
	\item $z_2=0$ and $z_1^\theta=j_1 z_1$, containing orbits on constant planes $\theta = const$.
	\item $z_1=0$ and $z_2^\dag=-j_2z_2$, describing spherical orbits.
	\item Finally, the the most general case $z_2^\dag-z_1^\theta+j_1z_1+j_2 z_2=0$.
\end{enumerate}
This exhaustive classification contains all possible cases for geodesic motions. In the following subsections, we shall present the corresponding equations for each alternative and later apply them to the particular case of Kerr space-time.

\subsection{ Circular orbit on a plane}
In the first case the system (\ref{M2G1})-(\ref{M3G2}) is solved by  $z_2=0$ and $z_1=0$, i.e the radial and the angular coordinate, $\theta$, are constant. Thus, we obtain bounded orbits confined to a plane and the set of equations becomes,
\begin{eqnarray}
  j_6z_3^2+2z_0z_3\Omega_2-a_1z_0^2 &=& 0 \, , \label{CircOrbit_1} \\
  j_9z_3^2+2z_0z_3\Omega_3-a_2z_0^2 &=& 0 \, , \label{CircOrbit_2} \\
  z_1=\frac{f_1^\prime}{B}&=&0 \label{CircOrbit_3} \, , \\
{\rm and} \quad z_2=\frac{f_{2,\theta}}{C}&=&0\, \label{CircOrbit_4}.
\end{eqnarray}

\subsection{General  orbital motion on a constant plane}
The second solution for equations (\ref{M2G1})-(\ref{M3G2}), emerges from the conditions $z_2=0$ and $z^{\theta}_1 = j_1 z_1$. Then, the geodesic equations reduce to
\begin{eqnarray}
  j_6z_3^2+ 2z_0z_3 \Omega_2-a_1z_0^2 - z_1 z_1^\dag &=&0 \, , \label{ecB11}\\
  j_9z_3^2+2z_0z_3\Omega_3-a_2z_0^2-j_1 z_1^2 &=& 0 \, , \label{ecB1}\\
  z_1=\frac{f_1^\prime}{B}&\neq &0 \quad {\rm and}\\
 z_2=\frac{f_{2,\theta}}{C}&=&0\label{ecB14}\, .
\end{eqnarray}

\subsection{General  orbits on the two-sphere}
The third set of solutions we shall consider are general orbits circumscribed on a 2-sphere ($\theta-\phi$), recently reported in reference~\cite{Teo2021}. This occurs having $z_1=0$ and $z_2^\dag=-j_2z_2$, and the corresponding geodesic equations are
\begin{eqnarray}
  j_6z_3^2+2z_0z_3\Omega_2-a_1z_0^2+j_2 z_2^2 &=& 0 \, ,\label{j631}\\
  j_9z_3^2+2z_0z_3\Omega_3-a_2z_0^2- z_2z_2^\theta &=&0 \, , \\
  z_1=\frac{f_1^\prime}{B}&= &0  \quad {\rm and} \label{z13}\\
 z_2=\frac{f_{2,\theta}}{C}&\neq&0 \, . \label{j633}
\end{eqnarray}

\subsection{The general case}
The last case emerges from $z_2^\dag -z_1^\theta +j_1z_1 +j_2 z_2=0$. This is the most general case, and the system of geodesic equations becomes
\begin{eqnarray}
       j_6z_3^2+ 2z_0z_3 \Omega_2-a_1z_0^2 +j_2 z_2^2 - z_1 z_1^\dag &=&0 \, , \\
  {\rm and} \quad j_9z_3^2+2z_0z_3\Omega_3-a_2z_0^2-j_1 z_1^2- z_2z_2^\theta &=&0 \, ;\\
{\rm with} \quad z_1=\frac{f_1^\prime}{B}&\neq &0\label{f14}\\
{\rm and} \quad z_2=\frac{f_{2,\theta}}{C}&\neq&0\label{f24} \, .
\end{eqnarray}

The solution of the equation
\begin{equation}
z_2^\dag-z_1^\theta+j_1z_1+j_2 z_2=0\label{ecgeodesica}
\end{equation}
is
\begin{equation}
z_1=f^\dag\quad {\rm and} \quad z_2=f^\theta \, ,
\label{solgeodesica}
\end{equation}
with $f=f(r,\theta)$ an arbitrary function of its arguments. Consequently,
\begin{equation}
z_1^\theta=j_1z_1\quad {\rm and} \quad z_2^\dag=-j_2z_2 \, , \label{z1z2}
\end{equation}
    which in turn allows us to transform (\ref{G1}) and (\ref{G2}) into
\begin{equation}
z_1 z_1^\dag =2z_0z_3 \Omega_2-a_1z_0^2 +j_2 z_2^2 +j_6z_3^2 \label{G11}
\end{equation}
and
\begin{equation}
z_2z_2^\theta =2z_0z_3\Omega_3-a_2z_0^2-j_1 z_1^2+j_9z_3^2 \label{G22} \, .
\end{equation}

\section{Symmetry and Geodesic Equations}
In this section we shall discuss the consequences on imposing symmetries, i.e. Killing vectors and tensors, on the source generating the geodesic equations.
\subsection{Killing Vectors}
From the Killing equation
\begin{equation}
\mathfrak{L}_{X}g_{\alpha\beta}=g_{\delta\beta}X^\delta_{,\beta}+g_{\beta \delta}X^\delta_{,\beta}+g_{\alpha\beta,\delta}X^\delta \nonumber \, ,
   \label{EQ:Killing}
\end{equation}
we can identify  temporal and axial Killing vectors as
\begin{equation}
  \tau^\alpha=(1,0,0,0)=\tau_0 V^\alpha\quad \Rightarrow \tau_0=A \quad {\rm and}
  \label{EQ:tempKilling}
\end{equation}
\begin{eqnarray}
\xi^\alpha &=&(0,0,0,1)=\xi_0 V^\alpha +\xi_3 S^\alpha \quad {\rm where} \label{EQ:AxialKilling}\\
\xi_0&=&-\frac{\omega_3}{A}\quad {\rm and} \quad \xi_3=\frac{\sqrt{\Delta _2}}{A}\, .\nonumber
\end{eqnarray}
These Killing vectors provide the conserved quantities: the energy $E~=~\tau^\alpha Z_\alpha$ and angular momentum  $l~=~\xi^\alpha Z_\alpha$. Thus,
\begin{equation}
    z_{0;\alpha} = -z_0 a_1 K_\alpha-a_2 z_0 L_\alpha \quad \Rightarrow
    z_0=-\frac{E}{A} \, ,
    \label{Zzero}
\end{equation}
and
\begin{equation}
    z_{3;\alpha} = -(2z_0 \Omega_2+j_6z_3) K_\alpha-(2z_0 \Omega_3+j_9z_3)L_\alpha \, .
    \label{Z3derivative}
\end{equation}
Where we have written $z_3$ as
\begin{equation}
    z_3=\frac{A^2 l +E \omega_3}{A\sqrt{\Delta_2}} \, .
    \label{Z3}
\end{equation}

Now the set of equations for the parallel transport of the vector $Z^\alpha$,
 can be written as
  \begin{eqnarray}
    \frac{{\rm d}t}{{\rm d}\lambda} &=& \frac{z_0}{A}+\frac{z_3 \omega_3}{A\sqrt{\Delta_2}}=\frac{\omega_3  l -R^2 E}{\Delta_2} \label{tpto}\, ,\\
    \frac{{\rm d}r}{{\rm d}\lambda} &=& \frac{z_1}{B} \, , \label{rpto}\\
    \frac{{\rm d}\theta}{{\rm d}\lambda} &=& \frac{z_2}{C} \quad {\rm and}\label{thpto}\\
      \frac{{\rm d}\phi}{{\rm d}\lambda} &=& \frac{z_3 A}{\sqrt{\Delta_2}} =\frac{A^2  l +\omega_3 E}{\Delta_2}\, .\label{phpto}
  \end{eqnarray}
or equivalently the set of trajectories (without using the afine parameter $\lambda$),
\begin{eqnarray}
  \dot{r} &=& \frac{{\rm d} r}{{\rm d}t}=\frac{z_1 \Delta_2}{B(\omega_3  l -R^2 E)}\label{geort} \\
  \dot{\theta} &=& \frac{{\rm d}\theta}{{\rm d}t}=\frac{z_2(A^2R^2+\omega^2_3)}{C(\omega_3  l -R^2 E)}\quad {\rm and} \label{geott} \\
  \dot{\phi} &=&  \frac{{\rm d}\phi}{{\rm d}t}=\frac{A^2  l +\omega_3 E}{\omega_3  l -R^2 E } \, . \label{geopt}
\end{eqnarray}
The above equations lead to the following characteristic expressions:
\begin{equation}
  \frac{B{\rm d}r}{z_1} = \frac{C {\rm d}\theta}{z_2}=\frac{\Delta_2 {\rm d}\phi}{A^2  l +\omega_3 E}\, . \label{geortp}
\end{equation}

\subsection{Constant azimuthal geodesic only for static cases}
A first general result emerges from equations (\ref{Z3derivative}) and  (\ref{phpto}) for $z_3=0$. It is clear that it only admits the solution $l=0$ in addition to $w_3=0$ since the other possibility $w_3=-{l A^2}/{E}$ must be rejected because of asymptotic conditions. Thus,  the metric functions $w_3$ and $A$  cannot be proportional through a constant since their asymptotic behaviours are not compatible; consequently {\bf constant azimuthal geodesics} are the only possibilities for static metrics.

\subsection{Killing Tensor}
Killing tensors are useful because they also provide conserved
quantities for geodesic motion. The most famous is obtained for the Kerr space-time where the
Killing tensor leads to the Carter constant \cite{BCarter1968}.

For the stationary axially symmetric space-time, the Killing tensor $\xi_{\alpha\beta}$  satisfies
\begin{equation}
\xi_{\alpha\beta;\mu}+\xi_{\mu\alpha;\beta}+\xi_{\beta\mu;\alpha}=0\label{KillingEq}
\end{equation}
which can be written as
\begin{eqnarray}\label{KillingTensor}
  \xi_{\alpha \beta}&=&\xi_{00}V_\alpha V_\beta+\xi_{11}K_\alpha K_\beta+\xi_{22}L_\alpha L_\beta\nonumber\\
  &+&\xi_{33}S_\alpha S_\beta+\xi_{03}(V_\alpha S_\beta+V_\beta S_\alpha)
\end{eqnarray}

Integrating the above Killing equation (\ref{KillingEq}) we obtain
\begin{eqnarray}
  \xi_{00} &=& \xi(r)+\frac{\omega_3^2}{A^2} \, , \\
   \xi_{11}&=& C^2+\xi(r) \, ,\\
    \xi_{22}&=&\xi(r)\, , \\
  \xi_{33}&=& \xi(r)+\frac{\Phi^2}{A^2} \quad {\rm and}\\
  \xi_{03} &=& \frac{\Phi \omega_3}{A^2} \, .
\end{eqnarray}
We have defined $\Phi=(r^2-2mr+a^2)\sin \theta$  and $\xi(r)$ an arbitrary $r$-function.

Thus, it  provides a general conserved quantity $Q$ asociated with $\xi_{\alpha \beta}$  along the geodesic as
\begin{equation}\label{ConservQ}
  Q=\xi_{\alpha\beta}Z^\alpha Z^\beta\quad \Rightarrow\quad  Q_{;\mu}Z^\mu=0 \, .
\end{equation}
Now, since $Z$ has a constant modulus we found
\begin{eqnarray}
  z_1^2+z_2^2 &=& \epsilon+z_0^2-z_3^2 \quad {\rm and} \label{ConservQ1} \\
  \xi_{11} z_1^2+\xi_{22} z_2^2 &=& Q+2\xi_{03} z_0z_3-\xi_{00}z_0^2-\xi_{33}z_3^2 \, ,
  \label{ConservQ2}
\end{eqnarray}
and thus obtaining that the scalars $z_1$ and $z_2$ are
\begin{equation}
    z_1=\frac{\sqrt{g_1(r,\theta)}}{C}\quad {\rm and} \quad  z_2=\frac{\sqrt{g_2(r,\theta)}}{C} \, ,
\end{equation}
 with
\begin{eqnarray}
g_1(r,\theta)&=&Q-\xi_{22}\epsilon+2\xi_{03} z_0z_3-(\xi_{00}+\xi_{22})z_0^2+ \nonumber \\
             & & \qquad +(\xi_{22}-\xi_{33})z_3^2 \quad {\rm and} \label{g1General} \\
g_2(r,\theta)&=&\xi_{11}\epsilon-Q -2\xi_{03} z_0z_3+(\xi_{00}+\xi_{11})z_0^2+ \nonumber \\
             & & \qquad  +(\xi_{33}-\xi_{11})z_3^2 \,. \label{g2General}
\end{eqnarray}
Notice that $Q$ is constant along the geodesic for an axially symmetric space-times. This new conserved quantity recovers the Carter constant $Q_c$ for the Kerr metric (\cite{Carter1968, BCarter1968})

In the following sections, we shall implement all the previous cases to the particular example of the Kerr space-times.

\section{The Kerr metric}
To illustrate the different cases mentioned above, we consider the Kerr metric, written as
 \begin{eqnarray}
      {\rm ds}^2&=&- \left(1-\frac{2mr}{r^2+a^2\cos ^2\theta}\right){\rm d}t^2-\frac{4mar\sin^2\theta}{r^2+a^2\cos^2\theta}{\rm d}t {\rm d}\phi\nonumber \\
      &+&\frac{r^2+a^2\cos^2\theta}{r^2-2mr+a^2}{\rm d}r^2+(r^2+a^2\cos^2\theta){\rm d}\theta^2\nonumber \\
      &+&\sin^2\theta \left(r^2+a^2+\frac{2ma^2r \sin^2\theta}{r^2+a^2\cos^2\theta}\right){\rm d}\phi^2 \, .
      \label{Kerrmetric}
    \end{eqnarray}

\subsection{Kerr killing tensor and geodesic motions}
 It is easy to verify that, assuming the Kerr metric (\ref{Kerrmetric}), the general solution of the system (\ref{M2G1})-(\ref{M3G2}), for $z_1~\neq~0$ and $z_2~\neq~0$ is
 \begin{equation}
z_1^\theta=j_1 z_1\quad {\rm and}  \quad z_2^\dag=-j_2z_2 \, .\label{j1j2}
\end{equation}
Thus, the separation constant method devised by Carter is equivalent to solve the geodesic equations (\ref{G11}) and (\ref{G22}), for the Kerr metric where equations (\ref{g1General}) and (\ref{g2General})  become $g_1(r,\theta)~=~g_1(r)$ and $g_2(r,\theta)~=~g_2(\theta)$. Thus, we obtain
\begin{equation*}
g_1(r)=Q_c+r^2\epsilon+\frac{E^2(r^4+(2mr+r^2)a^2)+4marEl+a^2l^2}{r^2-2mr+a^2}
\end{equation*}
and
\begin{equation*}
g_2(\theta)=-Q_c-\frac{l^2}{\sin^2\theta}+a^2(\epsilon+E^2)\cos^2\theta\, ,
\end{equation*}
where $Q_c$ is the Carter constant.

\subsection{The scalars of the Killing tensor }
The killing tensor for the Kerr metric is
\begin{eqnarray}
  \xi_{00} &=& -\frac{2ma^2 r \cos^2\theta}{r^2+a^2\cos^2\theta}-\frac{4m^2a^2 r^2 \sin^2\theta}{(r^2+a^2\cos^2\theta)(F^2-a^2\sin^2\theta)}\nonumber , \\
  \xi_{03} &=&\frac{2marF\sin^3\theta}{a^2\sin^2\theta-F^2}\nonumber \, ,\\
  \xi_{11} &=& a^2 \cos^2\theta \nonumber \, ,\\
  \xi_{22} &=& -r^2 \quad {\rm and} \nonumber\\
   \xi_{33} &=& \frac{F^2r^2+a^4\cos^2\theta \sin^2\theta}{a^2\sin^2\theta-F^2}\, ;\nonumber
\end{eqnarray}
 with $F^2=r^2+a^2-2mr$.

\subsection{ Kerr circular orbit on a plane}
Considering the set of equations (\ref{CircOrbit_1})-(\ref{CircOrbit_4}) for the  metric (\ref{Kerrmetric}) with $\theta=\frac{\pi}{2}$, $Q_c=l^2$   we get:
\begin{equation}
(\epsilon+E^2)r^3 - 2m\epsilon r^2+(a^2(\epsilon +E^2)-l^2)r + 2m(Ea+l)^2=0 \label{ec11}
\end{equation}
and
\begin{eqnarray}
&&(4\epsilon+5E^2)mr^3-(6m^2(2\epsilon+E^2)-a^2(\epsilon+E^2)+l^2)r^2\nonumber\\
&+&m(a^2(\epsilon +4E^2)+6aEl+8\epsilon m^2+2l^2)r\label{ec22}\\
&-&2m^2a^2(\epsilon-2E^2)-4aElm^2=0\nonumber\, .
\end{eqnarray}
Equations (\ref{ec11}) and (\ref{ec22}) determine the radius of the circumference and relate the physical constants. For instance, when we have the specific case $\epsilon=a=0$ we obtain that  $r=3m$ and $l=\sqrt{27}Em$.

\subsection{Kerr general  orbital motion on a constant plane}
Equations (\ref{ecB11})-(\ref{ecB14}) with $\theta=\frac{\pi}{2}$, and $Q=l^2$ lead to
\begin{equation}
    z_1^2=\frac{(\epsilon +E^2)r^3-2m\epsilon r^2+(a^2(\epsilon+E^2)-l^2)r+2m(aE+l)^2}{r^3-2mr^2+a^2r} .
\end{equation}

For a null geodesics and $a=-\frac{l}{E}$ we integrate (\ref{geortp}) as
\begin{equation}
    r=m+\sqrt{a^2-m^2}\tan (\frac{\sqrt{a^2-m^2}}{a}(\phi_0-\phi))
\end{equation}
where $\phi_0$ is a constant of integration.

Now for a time-like geodesic we have
\begin{equation}
    \frac{x_1(r)+x_2(r)}{1-x_1(r) x_2(r)}=\tan(\beta_0 (\phi-\phi_0)) \, ,
\end{equation}
with
\begin{eqnarray}
x_1(r)&=&\frac{\beta_1+\beta_2 r}{\beta_3\sqrt{(E^2-1)r^2+2mr-a^2}} \\
{\rm and} \quad x_2(r)&=&\frac{\beta_4+\beta_5 r}{\beta_6\sqrt{(E^2-1)r^2+2mr-a^2}} \, ;
\end{eqnarray}
where we have these functions of the physical parameters
\begin{eqnarray}
\beta_0&=&\frac{2E\sqrt{m^2-a^2}\sqrt{a^2-2m^2-2m\sqrt{m^2-a^2}}}{m+\sqrt{m^2-a^2}} \, ,\\
\beta_1&=&a^2-m^2+m\sqrt{m^2-a^2} \, ,\\
\beta_2&=&(1-E^2)(m-\sqrt{m^2-a^2})-m \, ,\\
\beta_3&=&E\sqrt{a^2-2m^2+2m\sqrt{m^2-a^2}} \, ,\\
\beta_4&=&a^2-m^2-m\sqrt{m^2-a^2}\, ,\\
\beta_5&=&(1-E^2)(m+\sqrt{m^2-a^2})-m \quad {\rm and}\\
\beta_6&=&E\sqrt{a^2-2m^2-2m\sqrt{m^2-a^2}} \, .
\end{eqnarray}

\subsection{Kerr general  orbits on the two-sphere}
In this case, from equations (\ref{z13})-(\ref{j633}) we get
\begin{equation}
 (\epsilon +E^2)r^3-2m\epsilon r^2+(a^2(\epsilon+E^2)-l^2)r+2m(aE+l)^2=0
\end{equation}
and
\begin{equation}
    f_{2,\theta}=\frac{\cos\theta \sqrt{\tilde{a}\sin^2\theta-l^2}}{\sin\theta} \, ; \label{f23}
\end{equation}
where we have redefined $\tilde{a}=a^2(\epsilon+E^2)$.
\begin{widetext}
Next, substituting (\ref{f23}) into (\ref{geortp}) we obtain
\begin{equation}
\frac{(F^2l-\tilde{E})}{F^2\sqrt{\tilde{a}-l^2}} \arctan\left [\sqrt{\frac{\tilde{a}\sin^2\theta-l^2}{\tilde{a}-l^2}}\right ] +
 \arctan\left [\frac{\sqrt{\tilde{a}\sin^2\theta-l^2}}{l}\right ]+\phi_0-\phi=0 \, ,
\end{equation}
with $\tilde{E}=2marlE+a^2l$ and $F^2=r^2-2mr+a^2$.


\subsection{The Kerr general case}
The module of $Z$ for the Kerr metric can be written as
\begin{equation}
(z_1^2+z_2^2)C^2 = (\epsilon +E^2)a^2\cos^2\theta-\frac{l^2}{\sin^2\theta}+\epsilon r^2\label{ecsep} +\frac{E^2(r^4+(r^2+2mr)a^2)+4marlE+a^2l^2}{r^2-2mr+a^2} \, .
\end{equation}
Substituting (\ref{solgeodesica}) into (\ref{ecsep}) we find
	\begin{equation}
	(r^2-2mr+a^2)(f^\prime)^2+(f_{,\theta})^2=
	(\epsilon +E^2)a^2\cos^2\theta-\frac{l^2}{\sin^2\theta}+\epsilon r^2 +\frac{E^2(r^4+(r^2+2mr)a^2)+4marlE+a^2l^2}{r^2-2mr+a^2} \, ,
	\end{equation}
which is an equation for the function $f$ with a solution $f=f_1(r)+f_2(\theta)$ where
	\begin{eqnarray}
	(f^\prime_1)^2&=&\frac{-Q+r^2\epsilon}{r^2-2mr+a^2}+\frac{E^2(r^4+(2mr+r^2)a^2)+4marEl+a^2l^2}{(r^2-2mr+a^2)^2}\label{f1p} \\
{\rm and} \quad	f^2_{2,\theta}&=&Q-\frac{l^2}{\sin^2\theta}+a^2(\epsilon+E^2)\cos^2\theta\label{f2t} \, .
	\end{eqnarray}
\end{widetext}

Next, substituting (\ref{f14}) and (\ref{f24})  into (\ref{geortp}) and considering (\ref{z1z2}), we obtain
\begin{eqnarray}
\frac{{\rm d}\theta}{f_{2,\theta}}  &=&  \frac{{\rm d}r}{f_1^\prime (r^2-2mr+a^2)} \quad {\rm and}
   \label{ec1Kerr}\\
{\rm d}\phi &=&  \left( \frac{(F^2-a^2\sin^2\theta)l-2marE\sin^2\theta}{F^4 f^\prime_1 \sin^2\theta}\right ){\rm d}r \, ,
  \label{ec2Kerr}
\end{eqnarray}
    where $F^2=r^2+a^2-2mr$.

Now, combining equations (\ref{ec1Kerr}) and (\ref{ec2Kerr}) we get
\begin{equation}
\frac{l {\rm d}\theta}{\sin^2 \theta f_{2,\theta}}-\frac{(a^2 l +2ma E r){\rm d}r}{F^4 f^\prime_1}={\rm d}\phi \, ,
\label{ec22Kerr}
\end{equation}
and by introducing
\begin{equation}
P(r)=a_0 r^4+a_1 r^3+a_2r^2+a_3r+a_4 \, ,
\label{Polinomy}
\end{equation}
equations (\ref{f1p}) and (\ref{f2t}) become
\begin{eqnarray}
  f_{2,\theta} &=& -\frac{\sqrt{P(\cos\theta)}}{\sin \theta} \quad {\rm and}\label{f22Kerr} \\
  \nonumber\\
   f_1^{\prime}&=& \frac{\sqrt{P(r)}}{r^2-2mr+a^2}\, .\label{f12Kerr}
\end{eqnarray}

Next, substituting $r=x+b$ into (\ref{Polinomy})  we get
\begin{equation}
\label{polf}
  P(x)=(\sqrt{a_0}x+b_1)^2(x+b_2)(x+b_3) \, ;
\end{equation}

with
\begin{eqnarray}
  a_4+a_3b+a_2b^2 +a_1b^3+a_0b^4 &=& b_1^2 b_2 b_3 \, ,  \\
    \nonumber \\
  a_3+2a_2b+3a_1b^2+4a_0b^3&=& b_1^2(b_2+b_3)+2\sqrt{a_0}b_1 b_2 b_3 , \nonumber \\
   \\
  a_2+3a_1b+6a_0b^2&=& (b_1+\sqrt{a_0}b_2)(b_1+\sqrt{a_0}b_3) \nonumber \\
   \\
{\rm and} \quad  a_1+4a_0b &=& 2\sqrt{a_0}b_1+a_0(b_2+b_3) \, .
\end{eqnarray}
\begin{widetext}
Consequently first integrals in $r$ and $\theta$  of the equation (\ref{ec22Kerr}) can be obtained as
\begin{equation}\label{InD}
\int \frac{(\kappa_1 x+\kappa_2)dx}{( x^2+s_1 x+s_2)\sqrt{(\sqrt{a_0} x+b_1)^2(x+b_2)(x+b_3)}}=\alpha_1 \arctan \gamma_1 \Gamma (x)+\alpha_2 \arctan \gamma_2  \Gamma (x)+\alpha_3 \arctan \gamma_3  \Gamma (x) \, ;
\end{equation}
where $ \Gamma  (x)=\sqrt{\frac{b_2+x}{b_3+x}}$,
with
\begin{eqnarray}
\gamma_1&=&\sqrt{\frac{b_1}{a_0b_2-b_1}} \, ,\qquad\qquad \qquad\qquad  \alpha_1=\frac{a_0(2b_1maE-a_0a^2l}{a^2a_0^2+b_1(b_1+2a_0m)} \, ,\\
\gamma_2&=&\sqrt{\frac{\sqrt{m^2-a^2}-m}{m-b_2-\sqrt{m^2-a^2}}} \, ,\qquad\qquad \alpha_2=\frac{2b_2maE-a_0a^2l-\alpha_0}{a^2a_0^2+b_1(b_1+2a_0m)} \, ,\\
\gamma_2&=&\sqrt{\frac{\sqrt{m^2-a^2}+m}{b_2-m-\sqrt{m^2-a^2}}}\, , \qquad\qquad \alpha_3=\frac{2b_2maE-a_0a^2l+\alpha_0}{a^2a_0^2+b_1(b_1+2a_0m)}\\
\nonumber\\
{\rm and} \quad \alpha_0&=&\frac{2a^3a_0mE+a_0ma^2l+b_1(2m^2aE+a^2l)}{\sqrt{m^2-a^2}} \, .
\end{eqnarray}
Now, implementing the procedure described above for the polynomial (\ref{polf}) we get
\begin{equation}
P(r)=(\epsilon+E^2)r^4-2m\epsilon r^3+(a^2(\epsilon+E^2)-l^2)r^2 +2m(aE+l)^2 r \, ,
\end{equation}
\end{widetext}
finding $b=0$, $b_3=0$, $a_0=\epsilon+E^2$,
\begin{eqnarray}
b_1^2+\frac{2m\epsilon}{\sqrt{\epsilon+E^2}}b_1+a^2(\epsilon+E^2)-l^2&=&0\\
{\rm and } \quad b_2- \frac{2m(aE+l)^2}{b_1^2}&=&0 \, .
\end{eqnarray}
Next, integrating the equations (\ref{ec1Kerr}) and (\ref{ec22Kerr}) we get
\begin{widetext}
\begin{equation}
    \arctan\left[\frac{\sqrt{\tilde{a}\sin^2\theta-l^2}}{\tilde{a}-l^2}\right]-\sqrt{\frac{\tilde{a}-l^2}{b_1(a_0b_2-b_1)}}\arctan\left[\sqrt{\frac{a_0b_2-b_1}{b_1}}\sqrt{\frac{r}{r+b_2}}\right]=C_1
\end{equation}
and
\begin{eqnarray}
   && \arctan\left[\frac{\sqrt{\tilde{a}\sin^2\theta-l^2}}{l}\right]+ \frac{l}{\sqrt{\tilde{a}-l^2}} \arctan\left[\sqrt{\frac{\tilde{a}\sin^2\theta-l^2}{l^2-\tilde{a}}}\right] -2\alpha_1 \arctan\left[\gamma_1 \sqrt{\frac{r}{b_2+r}}\right]\nonumber\\
    &&-\alpha_2 \arctan\left[\gamma_2 \sqrt{\frac{r}{b_2+r}}\right]-\alpha_3 \arctan\left [\gamma_3 \sqrt{\frac{r}{b_2+r}}\right ]=C_2
\end{eqnarray}
\end{widetext}
Thus, the general solution is an arbitrary function
\begin{equation}
    G=G(C_1,C_2)
\end{equation}

\section{Conclusions}

This work presents a method to classify and solve all geodesic motion analytically around any stationary axially symmetric source. The method summarises all these possible geodesic trajectories into two simple equations (\ref{M2G1}) and (\ref{M3G2}), with ease to obtain solutions. These distinct solutions allow us to classify the different trajectories for particles and photons. All the possible geodesics have been implemented for the Kerr metric, representing the gravitational field produced by a rotating compact object. In particular, those orbiting on a two-sphere surface could be especially relevant for describing observational data ( see \cite{AbuterEtal2018A, AbuterEtal2018B, AbuterEtal2020} and references therein). Now, it will be possible to build templates from exact General Relativistic analytical solutions, i.e. without any approximations for the orbits of stars~\cite{GhezEtal2008, GenzelEisenhauerGillessen2010, AbuterEtal2018A}, and the imaging of black holes~\cite{AkiyamaEtal2019EHTColl, PsaltisEtalEHT2020, VelasquezEtal2022}. The method presented here allows us to avoid elliptic integrals in writing down the geodesics trajectories ( see for example references \cite{FujitaHikida2009, HackmannEtal2010, Lammerzahl2016} ).

We found the most general for this Killing tensor corresponding to any axisymmetric space-time and its linked constant of motion. The existence this general constant of motion --along the geodesic-- is clear from a simple system of algebraic equations (\ref{ConservQ1}) and (\ref{ConservQ2}).  Again, the general expression for the constant along the geodesic could help to obtain solutions for the geodesic in a more general context where the Kerr metric may not adequately describe the gravitational field (see \cite{DestounisKokkotas2021} and references therein). This new conserved quantity recovers the Carter constant for the Kerr metric (\cite{Carter1968, BCarter1968}).

Although we have considered Kerr space-time a helpful example, the equations for each case in our classification are general and valid for any axisymmetric metric. Analytic solutions for geodesic with more complex Kerr-like sources describing richer rotational compact objects could fit better the trajectories of the stars or represent more accurate black hole imaging or open the possibility of new information from gravitational wave astronomy.

\begin{acknowledgments}

The authors thank Prof. Georgios O Papadopoulos for pointing us to the importance of the Killing tensor for a general axisymmetric space-time. LAN gratefully acknowledges the support of the Vicerrector\'ia de Investigaci\'on y Extensi\'on from Universidad Industrial de Santander under project VIE2814. This work was partially supported by Ministerio de Ciencia, Innovacion y Universidades. Grant number: PGC2018 096038 B I00, and Junta de Castilla y Leon. Grant number: SA096P20.
\end{acknowledgments}


\begin{thebibliography}{10}

\bibitem{GhezEtal2008}
A.~M Ghez, S.~Salim, N.N. Weinberg, and et~al.
\newblock Measuring distance and properties of the milky way's central
  supermassive black hole with stellar orbits.
\newblock {\em The Astrophysical Journal}, 689(2):1044, 2008.

\bibitem{GenzelEisenhauerGillessen2010}
R.~Genzel, F.~Eisenhauer, and S.~Gillessen.
\newblock The galactic center massive black hole and nuclear star cluster.
\newblock {\em Reviews of Modern Physics}, 82(4):3121, 2010.

\bibitem{AbuterEtal2018A}
R.~{Abuter}, A.~{Amorim}, M~{Baub{\"o}ck}, et~al, and {Gravity Collaboration}.
\newblock Detection of orbital motions near the last stable circular orbit of
  the massive black hole sgra.
\newblock {\em Astronomy \& Astrophysics}, 618:L10, 2018.

\bibitem{AkiyamaEtal2019EHTColl}
K.~Akiyama, A.~Alberdi, R.~Azulay, et~al, and EHT Collaboration.
\newblock First m87 event horizon telescope results. i. the shadow of the
  supermassive black hole.
\newblock {\em The Astrophysical Journal Letters}, 875(1):L4, 2019.

\bibitem{PsaltisEtalEHT2020}
D.~Psaltis, L.~Medeiros, P.~Christian, et~al, and EHT Collaboration.
\newblock Gravitational test beyond the first post-newtonian order with the
  shadow of the m87 black hole.
\newblock {\em Physical review letters}, 125(14):141104, 2020.

\bibitem{HeLin2021}
G.~He and W.~Lin.
\newblock Kerr-newman black hole lensing of relativistic massive particles in
  the weak field limit.
\newblock {\em arXiv preprint arXiv:2112.08142}, 2021.

\bibitem{CostaNatario2021}
L.F. Costa and J.~Nat{\'a}rio.
\newblock Frame-dragging: meaning, myths, and misconceptions.
\newblock {\em Universe}, 7(10):388, 2021.

\bibitem{YuanNarayan2014}
F.~Yuan and R.~Narayan.
\newblock Hot accretion flows around black holes.
\newblock {\em Annual Review of Astronomy and Astrophysics}, 52:529--588, 2014.

\bibitem{AbbottEtall2016LIGOVIRGOColl}
B.P. Abbott, R.~Abbott, T.D. Abbott, et~al, and LIGO-Virgo Collaboration.
\newblock Observation of gravitational waves from a binary black hole merger.
\newblock {\em Physical review letters}, 116(6):061102, 2016.

\bibitem{Kerr1963}
R.P. Kerr.
\newblock Gravitational field of a spinning mass as an example of algebraically
  special metrics.
\newblock {\em Physical review letters}, 11(5):237, 1963.

\bibitem{Carter1968}
B.~Carter.
\newblock Global structure of the kerr family of gravitational fields.
\newblock {\em Physical Review}, 174(5):1559, 1968.

\bibitem{BCarter1968}
B.~Carter.
\newblock Hamilton-jacobi and schrodinger separable solutions of einstein's
  equations.
\newblock {\em Commun. Math. Phys.}, 10:280, 1968.

\bibitem{Teo2021}
E.~Teo.
\newblock Spherical orbits around a kerr black hole.
\newblock {\em General Relativity and Gravitation}, 53(1):1--32, 2021.

\bibitem{ChanEtal2018}
C.~Chan, L.~Medeiros, F.~{\"O}zel, and D.~Psaltis.
\newblock Gray2: a general purpose geodesic integrator for kerr spacetimes.
\newblock {\em The Astrophysical Journal}, 867(1):59, 2018.

\bibitem{OspinoHernandezNunez2017}
J.~{Ospino}, J.~L. {Hern{\'a}ndez-Pastora}, and L.~A. {N{\'u}{\~n}ez}.
\newblock An equivalent system of einstein equations.
\newblock {\em Journal of Physics Conference Series}, 831:012011, March 2017.

\bibitem{OspinoEtal2018}
J.~Ospino, J.L. Hern{\'a}ndez-Pastora, H.~Hern{\'a}ndez, and L.A.
  N{\'u}{\~n}ez.
\newblock Are there any models with homogeneous energy density?
\newblock {\em General Relativity and Gravitation}, 50(11):146, 2018.

\bibitem{OspinoNunez2020}
J.~{Ospino} and L.~A. {N\'u\~nez}.
\newblock Karmarkar scalar condition.
\newblock {\em The European Physical Journal C}, page 166, January 2020.

\bibitem{AbuterEtal2018B}
R.~{Abuter}, A.~{Amorim}, N.~{Anugu}, et~al, and {Gravity Collaboration}.
\newblock {Detection of the gravitational redshift in the orbit of the star S2
  near the Galactic centre massive black hole}.
\newblock {\em Astronomy \& Astrophysics}, 615:L15, July 2018.

\bibitem{AbuterEtal2020}
R.~{Abuter}, A.~{Amorim}, et~al, and {Gravity Collaboration}.
\newblock {Detection of the Schwarzschild precession in the orbit of the star
  S2 near the Galactic centre massive black hole}.
\newblock {\em Astronomy \& Astrophysics}, 636:L5, April 2020.

\bibitem{VelasquezEtal2022}
J.M. Vel{\'a}squez-Cadavid, J.A. Arrieta-Villamizar, F.D. Lora-Clavijo, O.M.
  Pimentel, and J.E. Osorio-Vargas.
\newblock Osiris: a new code for ray tracing around compact objects.
\newblock {\em The European Physical Journal C}, 82(2):1--12, 2022.

\bibitem{FujitaHikida2009}
R.~Fujita and W.~Hikida.
\newblock Analytical solutions of bound timelike geodesic orbits in kerr
  spacetime.
\newblock {\em Classical and Quantum Gravity}, 26(13):135002, 2009.

\bibitem{HackmannEtal2010}
E.~Hackmann, C.~L{\"a}mmerzahl, V.~Kagramanova, and J.~Kunz.
\newblock Analytical solution of the geodesic equation in kerr-(anti-) de
  sitter space-times.
\newblock {\em Physical Review D}, 81(4):044020, 2010.

\bibitem{Lammerzahl2016}
E.~{L\"ammerzahl}, C.and~{ Hackmann}.
\newblock Analytical solutions for geodesic equation in black hole spacetimes.
  in: Nicolini p., kaminski m., mureika j., bleicher m. (eds) 1st karl
  schwarzschild meeting on gravitational physics.
\newblock {\em Springer Proceedings in Physics}, 2016.

\bibitem{DestounisKokkotas2021}
K.~Destounis and K.D. Kokkotas.
\newblock Gravitational-wave glitches: Resonant islands and frequency jumps in
  nonintegrable extreme-mass-ratio inspirals.
\newblock {\em Physical Review D}, 104(6):064023, 2021.

\end{thebibliography}

\end{document}